\documentclass[reprint,onecolumn,amsmath,amssymb,aps,]{revtex4-2}

\pdfoutput=1
\usepackage[utf8]{inputenc}

\usepackage{graphicx}
\usepackage{dcolumn}
\usepackage{bm}
\usepackage{mathtools}
\usepackage{mathrsfs}

\usepackage[]{hyperref}
\usepackage{color} 
\usepackage{amsmath}

\begin{document}
	
\title{Entanglement preservation and Clauser–Horne nonlocality in electromagnetically induced transparency quantum memories}
	
\author{Po-Han Tseng$^{1}$ and Yong-Fan Chen$^{1,2}$}
	
\email{yfchen@mail.ncku.edu.tw}
	
\affiliation{
$^1$Department of Physics, National Cheng Kung University, Tainan 70101, Taiwan\\ 
$^2$Center for Quantum Frontiers of Research $\&$ Technology, Tainan 70101, Taiwan	
}
	

%
\date{November 17, 2026}


\begin{abstract} 

Understanding how decoherence-induced loss affects entanglement and nonlocal correlations in quantum memories is important for quantum information science. Here we develop a quantitative description of photonic quantum states retrieved from electromagnetically induced transparency (EIT) memories in a ground-state-decoherence-limited regime. Using the dark-state-polariton formalism, we derive the attenuation factor of the retrieved probe mode from the EIT dynamics. This attenuation depends on the coupling field and ground-state coherence decay and is not introduced as an independent phenomenological loss parameter. Within a specified quantum-limited pure-loss model, we use this attenuation to obtain the reduced photonic density operator. We show that, in the symmetric case, the complete retrieved state of an initially path-entangled single photon remains entangled for any nonzero storage efficiency, despite acquiring a vacuum component. In contrast, violation of the unconditional Clauser--Horne inequality under displaced on/off detection without post-selection requires an intrinsic efficiency above 89.7\%. We further derive the retrieved $N$-qubit photonic density operator for multiple independently operating EIT memories. These results establish a quantitative connection between EIT memory dynamics, retrieved quantum states, and operational nonlocality benchmarks under explicit channel assumptions.

\end{abstract}


\maketitle


\newcommand{\FigOne}{
    \begin{figure}[t]
    \centering
    \includegraphics[width = 10.0 cm]{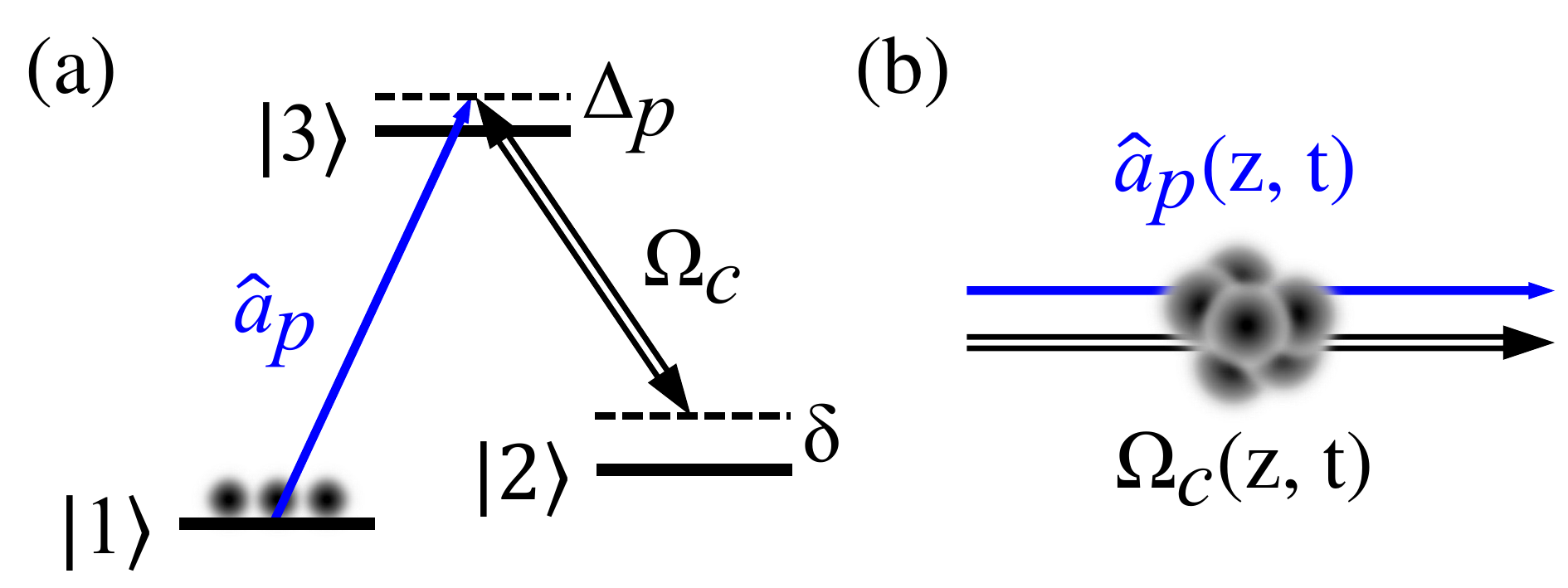}
    \caption{
EIT quantum memory in an atomic ensemble. (a) Energy-level diagram illustrating the $\Lambda$-type EIT scheme and the corresponding transitions for the two participating fields. (b) Schematic representation showing the propagation directions of the two participating fields, where all light fields propagate in the same direction.
}
    \label{fig:energy level diagram}
    \end{figure}
}

\newcommand{\FigTwo}{
    \begin{figure}[t]
    \centering
    \includegraphics[width = 10.0 cm]{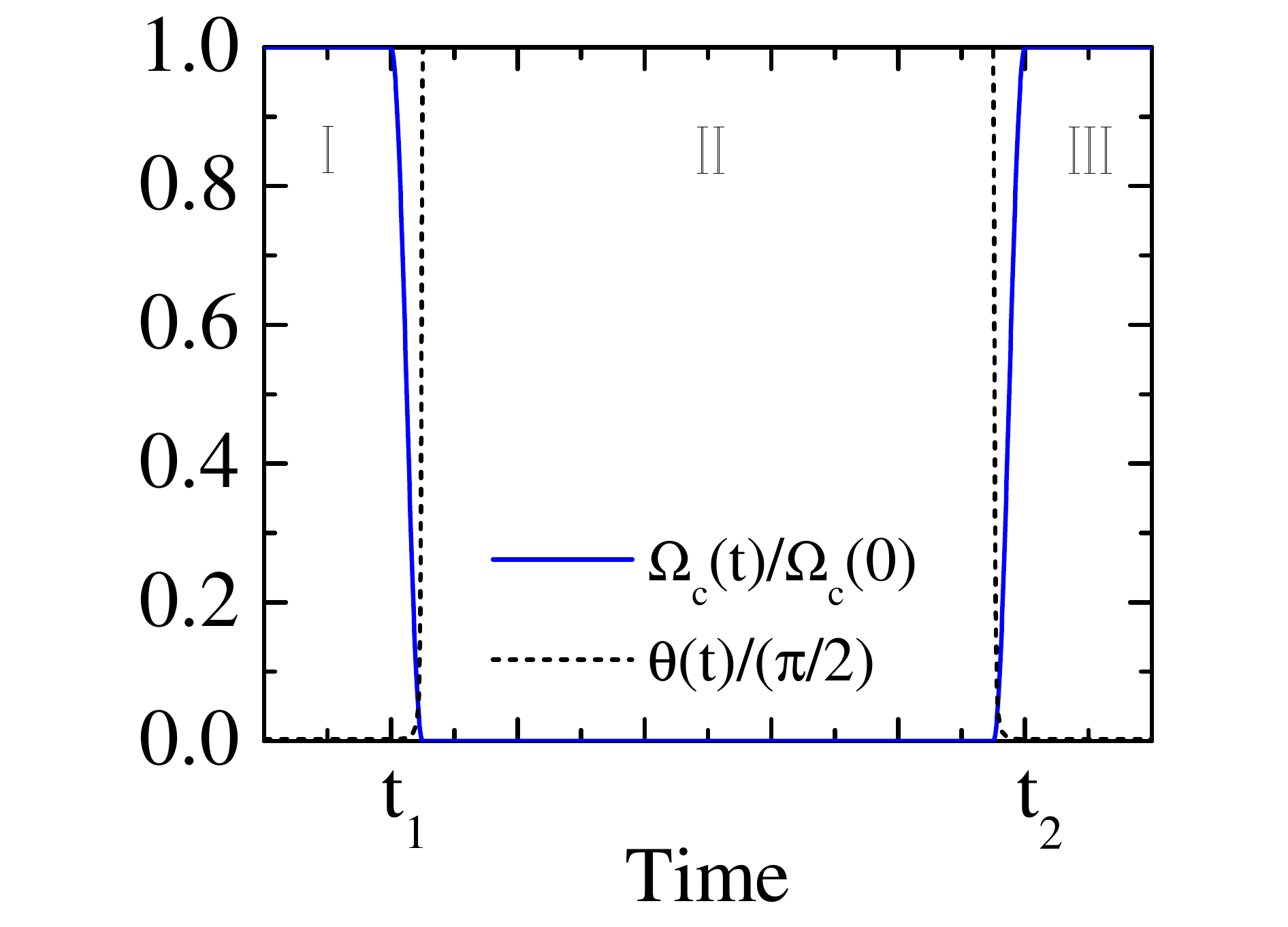}
    \caption{
The Rabi frequency of the coupling field $\Omega_c(t)$ and the mixing angle $\theta(t)$ as functions of time during the storage and retrieval processes in EIT quantum memory. For $t \leq t_1$ or $t \geq t_2$, the coupling Rabi frequency remains constant, and the mixing angle approaches zero, i.e., $\Omega_c(t) = \Omega_c(0)$ and $\theta(t) \approx 0$. During the storage interval ($t_1 \leq t \leq t_2$), except for the brief switching periods of the coupling field, $\Omega_c(t) = 0$ and $\theta(t) = \frac{\pi}{2}$. In this regime, the probe field is completely mapped into the spin-wave coherence of the atomic medium.
}
    \label{fig:DSP propagation}
    \end{figure}
}

\newcommand{\FigThree}{
    \begin{figure}[t]
    \centering
    \includegraphics[width = 15.0 cm]{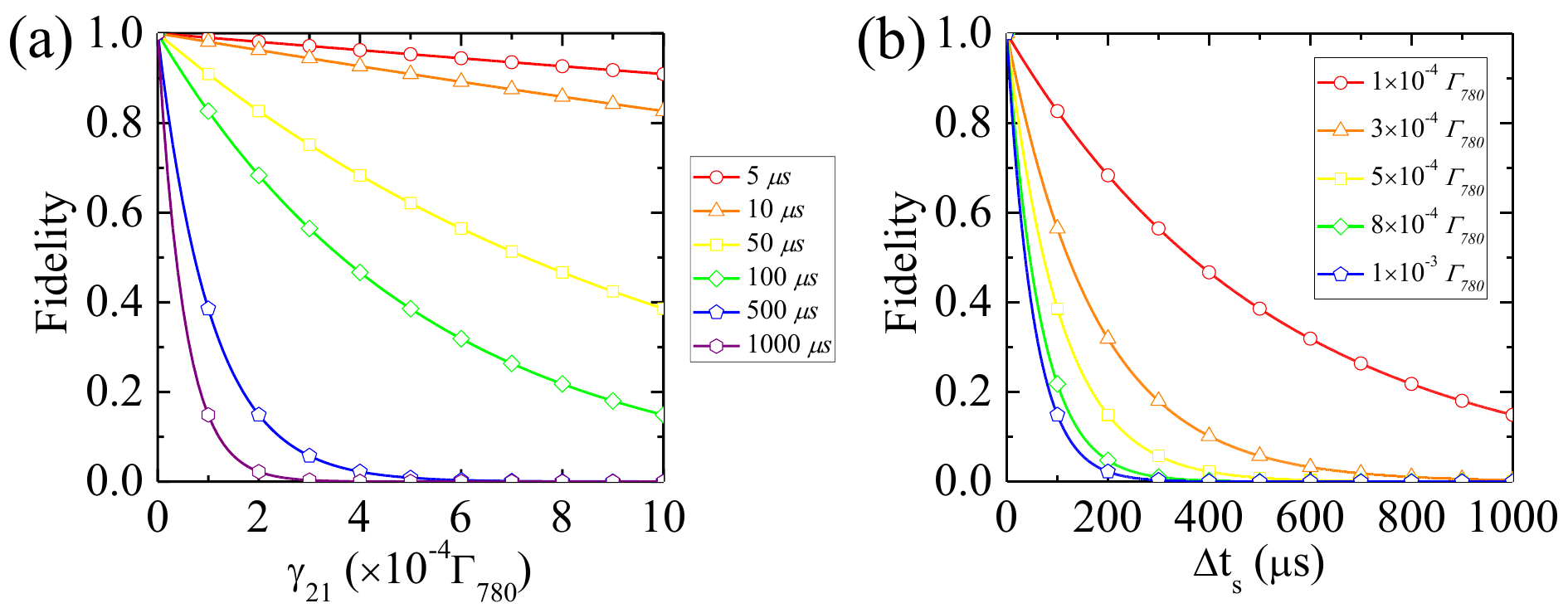}
    \caption{
Fidelity of EIT quantum memory as functions of ground-state decoherence rate $\gamma_{21}$ and storage time $\Delta t_s$. (a) Fidelity versus $\gamma_{21}$ for various $\Delta t_s$. (b) Fidelity versus $\Delta t_s$ for various $\gamma_{21}$. Here, $\Gamma_{780} = 2\pi \times 6.063$ MHz denotes the spontaneous decay rate of the $5P_{3/2}$ excited state of $^{87}\text{Rb}$. 
}
    \label{fig:multimode Fock state fidelity}
    \end{figure}
}

\newcommand{\FigFour}{
    \begin{figure}[t]
    \centering
    \includegraphics[width = 11.0 cm]{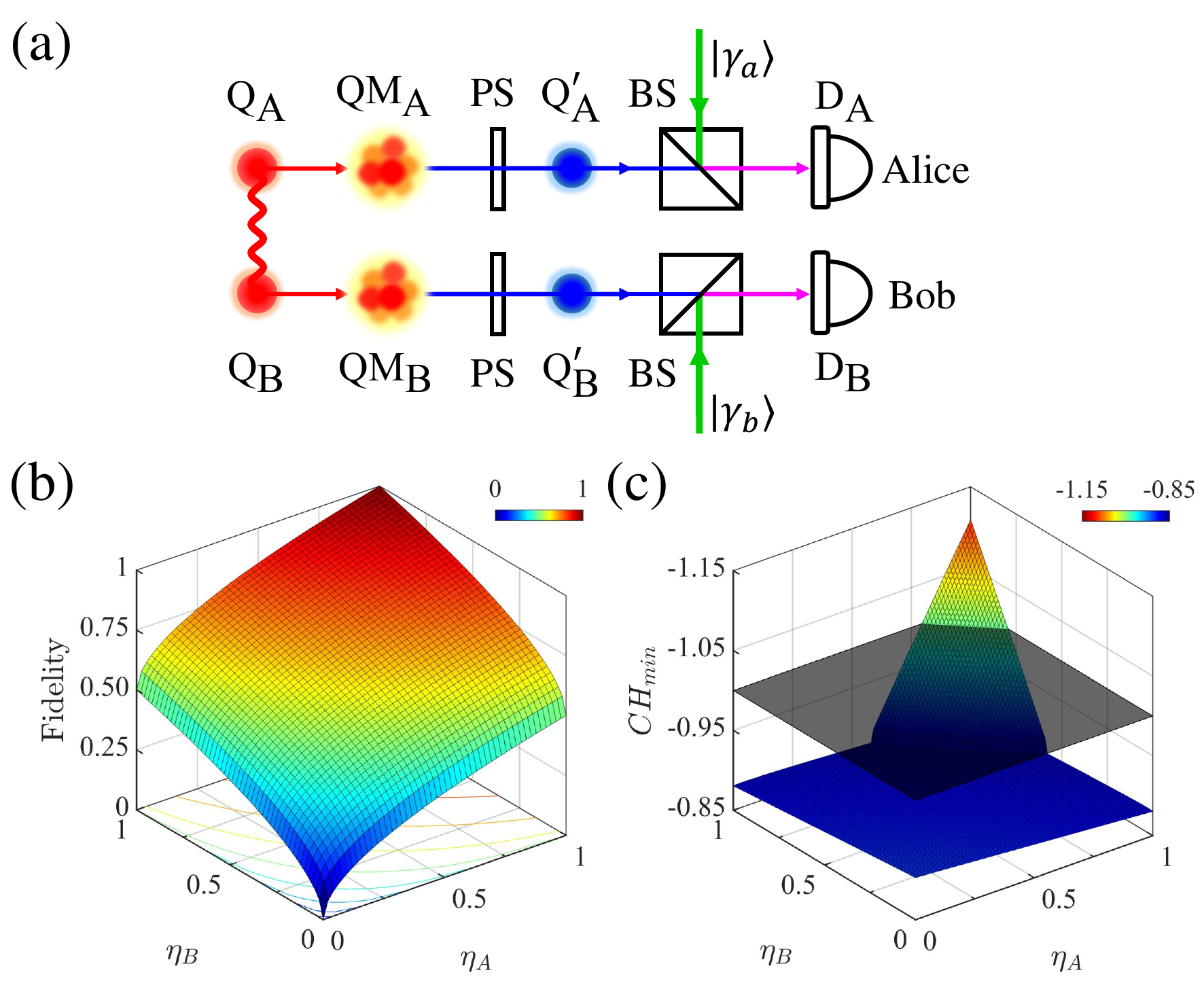}
    \caption{
Entanglement retention and an unconditional CH benchmark for nonlocality in EIT quantum memories. (a) Schematic of an EIT memory scheme for a pair of single-rail-encoded qubits $\mathrm{Q_A}$ and $\mathrm{Q_B}$. The two spatially separated qubits are stored simultaneously in respective EIT quantum memories, $\mathrm{QM_A}$ and $\mathrm{QM_B}$, with the phases of the retrieved fields corrected by phase shifters (PSs). In the CH Bell test, the retrieved qubits $\mathrm{Q_A'}$ and $\mathrm{Q_B'}$ are sent to Alice and Bob, respectively, after being combined with local oscillators $|\gamma_a\rangle$ and $|\gamma_b\rangle$ via beam splitters (BSs). (b) Fidelity between the input state $|\phi_p(0)\rangle$ and the retrieved state $\rho_p$ as a function of the storage efficiencies $\eta_A$ and $\eta_B$ in the EIT memory scheme. (c) Minimum value of the CH combination for the retrieved state $\rho_p$ as a function of the storage efficiencies $\eta_A$ and $\eta_B$. The dark plane indicates the CH inequality boundary ($CH = -1$) for local theories, with unconditional CH violation obtained in the region where $CH_{\text{min}} < -1$ under the specified no-post-selection measurement protocol.
}
    \label{fig:CH inequality}
    \end{figure}
}


\section{Introduction} \label{sec:introduction}

Preserving quantum entanglement in realistic, noisy environments represents a central challenge in quantum information science. Quantum memories, which enable the reversible mapping between photonic and atomic excitations, play a crucial role in addressing this challenge and underpin key technologies such as quantum repeaters, deterministic photon sources, and linear optical quantum computing \cite{Optical quantum memory-1, Optical quantum memory Qubit Applications, Optical quantum memory-2, Optical quantum memory-3, Optical quantum memory and application, Integrated quantum memory}. Among these applications, quantum repeaters use quantum memories to distribute and synchronize entanglement over long distances, as exemplified by the Duan–Lukin–Cirac–Zoller (DLCZ) protocol, where stored quantum states await successful entanglement purification before advancing the network \cite{Optical quantum memory and application, DLCZ Protocol, Quantum Repeaters, Quantum memory and quantum repeater-1, Quantum memory and quantum repeater-2, Quantum memory and quantum repeater imperfect memories, Quantum memory and quantum repeater experiment-1, Quantum memory and quantum repeater experiment-2, Quantum memory and quantum repeater experiment-3}. Deterministic single- and multi-photon sources similarly rely on quantum memories for on-demand photon retrieval and synchronization \cite{Single-Photon Sources-1, Single-Photon Sources-2, Deterministic single-photon source-1, Deterministic single-photon source-2, Deterministic single-photon source-3, Deterministic multi-photon source-1, Deterministic multi-photon source-2}. In linear optical quantum computing, quantum memories mitigate the probabilistic nature of gate operations and enable scalable optical architectures through temporal synchronization \cite{Linear Optical Quantum Computing-1, LOQC quantum memory}. These examples highlight that the development of quantum technologies critically relies on efficient and high-fidelity storage and retrieval of quantum states \cite{Maximally Path-Entangled Single-Photon Quantum Memory Atomic Ensemble, Polarization Encoded Qubits Quantum Memory Single Atom, Polarization Encoded Qubits Telecom Quantum Memory, Polarization Encoded Qubits Quantum Memory Atomic Ensemble}.

A variety of physical platforms have been developed to realize quantum memories, each offering distinct advantages and limitations. Rare-earth-ion-doped (REI-doped) solids provide excellent physical stability and compatibility with micro- and nano-fabrication techniques \cite{REI-doped solid review-1,REI-doped solid review-2,REI-doped solid review-3}. Representative systems include Pr:YSO, Er:YSO, Eu:YSO, Ti:Tm:LiNbO$_3$, and Nd:YVO crystals, which have been extensively studied as solid-state quantum memories \cite{PrYSO-1,PrYSO-2,PrYSO-3,PrYSO-4,PrYSO-5,PrYSO-6,PrYSO-7,ErYSO-2,EuYSO-1,EuYSO-2,EuYSO-3,EuYSO-4,TiTmLiNbO-1,TiTmLiNbO-2,NdYVO-1,NdYVO-2,NdYVO-3,NdYVO-4,NdYVO-5,NdYVO-6,NdYVO-7,NdYVO-8,NdYVO-9}. Despite remaining material and spectral challenges, these systems remain strong candidates for scalable quantum memory architectures \cite{REI-doped solid review-2}. Solid-state defect systems also provide promising memory platforms. Diamond defect centers, including nitrogen-vacancy and group-IV split-vacancy centers, offer long coherence times and compatibility with integrated quantum technologies \cite{Quantum memory NV centre-1,Quantum memory NV centre-2,Quantum memory NV centre-3,Quantum memory NV centre-4,Quantum memory NV centre-5,Quantum memory XV centre,Quantum memory Si-V centre-1,Quantum memory Si-V centre-2,Quantum memory Ge-V centre}. Semiconductor quantum-dot memories further provide strong compatibility with existing photonic and spin-based devices \cite{Quantum dot review,Quantum dot-1,Quantum dot-2,Quantum dot-3}. For atomic-based quantum communication and computation, both single optically trapped atoms \cite{Single atom review,Single atom-1,Single atom-2,Single atom-3} and alkali-metal atomic ensembles \cite{EIT Quantum Memory,Atomic ensemble review-1,Atomic ensemble review-2,Atomic ensemble review-3} have demonstrated significant potential. In particular, atomic ensembles, realized either in warm vapor cells \cite{Vapor Cell Quantum Memory Single-Photon Storage,Warm alkaline vapour-1,Warm alkaline vapour-2,Warm alkaline vapour-3} or laser-cooled atomic gases \cite{Telecom QFC Quantum Network Link Atomic Ensembles,Quantum Memory sub-Second Scale,Polarization Encoded Qubits Single Photon Quantum Memory,Cold atomic ensemble-1,Cold atomic ensemble-2}, provide an experimentally accessible route toward scalable quantum memory protocols.

Among these platforms, quantum memories based on $\Lambda$-type electromagnetically induced transparency (EIT) in cold atomic ensembles have emerged as a particularly versatile and well-established system \cite{Telecom QFC Quantum Network Link Atomic Ensembles, Light Storage Minute Scale, Quantum Memory sub-Second Scale, Polarization Encoded Qubits Single Photon Quantum Memory, Cold atomic ensemble-1, Cold atomic ensemble-2, Cold atomic ensemble-3}. Utilizing the $D_1$ (795 nm) and $D_2$ (780 nm) transitions in rubidium atoms, such systems are naturally compatible with both the DLCZ protocol and neutral-atom quantum logic platforms \cite{Neutral Atom Quantum Computing, Neutral Atom Quantum Simulation-1, Neutral Atom Quantum Simulation-2, Neutral Atom Quantum Algorithms, Neutral Atom High-Fidelity Entangling Gates, Phase-Stabled DLCZ Protocol, DLCZ Protocol Realization}. The dark-state-polariton (DSP) framework describes the adiabatic conversion between a propagating probe excitation and collective ground-state coherence during EIT storage and retrieval \cite{DSP-1,DSP-2,DSP two-photon linewidth,DSP decoherence,DSP collapses and revivals,DSP single- and double-Lambda media}. Earlier studies have also examined decoherence effects in collective photon memories \cite{DSP decoherence}. Building on these developments, we investigate how the EIT-dependent attenuation of a retrieved probe mode is reflected in its photonic density operator and in an unconditional Bell-test benchmark.

To address this question, we combine DSP propagation dynamics \cite{DSP-1,DSP-2,DSP decoherence} with the reduced density-operator formalism \cite{Reduce Density Matrix-1, Reduce Density Matrix-2, Reduce Density Matrix-3}. The coupling-field evolution and ground-state coherence decay determine the retrieved-mode attenuation factor $f(t)$. We then evaluate the corresponding photonic quantum state in a quantum-limited regime characterized by an initially uncorrelated vacuum reservoir, for which the input--output transformation takes the pure-loss-channel form. This connects the EIT response to state-level quantities, including the vacuum contribution, surviving single-photon coherence, and unconditional CH correlations. The specified regime isolates the role of intrinsic attenuation; additional noise and mode imperfections can be treated in extensions of the model.

Our analysis shows that the retrieved photonic state violates the Clauser--Horne (CH) inequality above a protocol-dependent storage-efficiency benchmark, indicating that unconditional CH nonlocality can be observed under the specified displaced on/off detection protocol without post-selection. Extending the framework to an array of $N$ spatially separated EIT memories, we further derive the retrieved $N$-qubit photonic density operator under independent ground-state-decoherence-induced loss and show that the input state is recovered in the ideal noiseless-memory limit.

The remainder of this paper is organized as follows. Section~\ref{sec:field operator} introduces the DSP description and determines the retrieved-mode attenuation within the Heisenberg--Langevin approach. Section~\ref{sec:retrieved quantum state} develops the corresponding reduced photonic density operator under the specified vacuum-environment conditions. Section~\ref{sec:entanglement retention} examines independent EIT memories and the unconditional CH benchmark for a path-entangled single photon. Section~\ref{sec:conclusion} summarizes the main results and their physical scope, with supporting calculations in the Appendices.


\FigOne

\section{Field Operator} \label{sec:field operator}

\subsection{Heisenberg-Langevin Approach} \label{subsec:HLE and MSE}

We consider a cold atomic ensemble consisting of $\Lambda$-type three-level atoms with two metastable ground states and one excited state, as depicted in Fig. \ref{fig:energy level diagram}(a). The strong coupling field is treated classically, while the weak probe field is quantized. The interaction strength between the coupling field and the electric dipole moment is described by the Rabi frequency $\Omega_c = 2d_{32}E_c/\hbar$, where $d_{32}$ represents the electric dipole matrix element. The coupling constant between the quantized probe field and the electric dipole moment is denoted as $g_p = d_{31}\epsilon_p/\hbar$, where $\epsilon_p = \sqrt{\frac{\hbar\omega_p}{2\epsilon_0 V}}$ is the amplitude of the probe field. The two participating light fields are multimode fields, assumed to propagate in the same direction, as shown in Fig. \ref{fig:energy level diagram}(b). All detunings between the fields and atomic resonances, denoted as $\Delta_p$, $\Delta_c$, and $\delta$, are included in the theoretical model. The $\Lambda$-type system is collectively described using the collective atomic operator approach \cite{DSP-2}. Under the rotating-wave approximation and the slowly varying amplitude (SVA) \cite{SVA}, the Hamiltonian of the entire system is expressed as follows:
\begin{align}
	\hat{H}_S=&-\frac{N\hbar}{2L}\int_0^L \big[ 2g_p\hat{a}_p(z,t)\hat{\sigma}_{31}(z,t)+\Delta_p\hat{\sigma}_{33}(z,t)\notag\\
	&+\Omega_c(z,t)\hat{\sigma}_{32}(z,t)+\delta\hat{\sigma}_{22}(z,t)+h.c.\big] dz, \label{eq:(1)}
\end{align}
where $N$ represents the total number of atoms, $L$ denotes the length of the atomic ensemble, and $\hat{\sigma}_{ij}(z,t)$ is the collective atomic operator under the SVA. Phase mismatch is not a concern for the $\Lambda$-type system. The evolution of $\hat{\sigma}_{ij}(z,t)$ is described by the following Heisenberg-Langevin equations (HLEs) \cite{Quantum Optics}:
\begin{align}
	\frac{\partial\hat{\sigma}_{ij}}{\partial t}=\frac{i}{\hbar}\big[\hat{H}_S,\hat{\sigma}_{ij}\big]+\hat{R}_{ij}+\hat{F}_{ij}, \label{eq:(2)}
\end{align}
where $\hat{R}_{ij}$ denotes the relaxation term and $\hat{F}_{ij}$ the corresponding Langevin noise operator. Following the standard quantum-optical Heisenberg--Langevin approach \cite{Quantum Optics,Diamond-type QFC}, we describe the coherent atom--field interaction through $\hat{H}_{S}$ and include atomic relaxation through effective Markovian damping and noise terms. The ground-state coherence-decay parameter sets the attenuation studied below; the vacuum-environment conditions used to determine the reduced optical state are specified in Sec.~\ref{sec:retrieved quantum state}.

For the $\Lambda$-type system, the weak probe field is treated as a perturbation. The zeroth-order HLEs are solved under the steady-state assumption for zeroth-order populations. The HLEs for $\hat{\sigma}_{11}$, $\hat{\sigma}_{22}$, $\hat{\sigma}_{23}$, $\hat{\sigma}_{32}$, and $\hat{\sigma}_{33}$ are decoupled from the others, and the zeroth-order solutions to these decoupled equations are expressed as follows:
\begin{align}
	\hat{\sigma}_{ij}^{(0)}(z)=\langle\hat{\sigma}_{ij}^{(0)}(z)\rangle+\sum_{kl}\epsilon_{kl}(z)\hat{F}_{kl}(z). \label{eq:(3)}
\end{align}
The expectation values of the zeroth-order collective atomic operators are all zero except for $\langle\hat{\sigma}_{11}^{(0)}(z)\rangle = 1$. To determine the first-order atomic operators, the zeroth-order results are substituted into the corresponding first-order HLEs as follows:
\begin{gather}
	\frac{\partial}{\partial t}\hat{\sigma}_{13}^{(1)}=i\hat{a}_p g_p+\frac{i}{2}\hat{\sigma}_{12}^{(1)}\Omega_c-\frac{1}{2}\gamma_{31}'\hat{\sigma}_{13}^{(1)}+\hat{F}_{13},\label{eq:(4)}\\
	\frac{\partial}{\partial t}\hat{\sigma}_{12}^{(1)}=\frac{i}{2}\Omega_c^*\hat{\sigma}_{13}^{(1)}-\frac{1}{2}\gamma_{21}'\hat{\sigma}_{12}^{(1)}+\hat{F}_{12},\label{eq:(5)}
\end{gather}
where $\gamma_{21}' = \gamma_{21} - 2i\delta$, $\gamma_{31}' = \gamma_{31} - 2i\Delta_p$, and $\gamma_{ij}$ denotes the coherence-decay rate between states $|i\rangle$ and $|j\rangle$. In particular, $\gamma_{21}$ characterizes the decay of ground-state coherence in the linearized EIT dynamics. Products of the weak probe operator with Langevin noise operators, such as $\hat{a}_{p}\hat{F}_{ij}$, are neglected at the adopted perturbative order, while the Langevin-driven contribution to the retrieved field is accounted for in the effective output-channel description below.

To study the evolution of the probe field propagating through the $\Lambda$-type atomic medium, we employ the Maxwell-Schr\"{o}dinger equation (MSE):
\begin{align}
	\left(\frac{1}{c}\frac{\partial}{\partial t}+\frac{\partial}{\partial z}\right)\hat{a}_p(z,t)&=i\frac{N}{c}g_p^*\hat{\sigma}_{13}^{(1)}(z,t).\label{eq:(6)}
\end{align}
The absorption loss of the coupling field is neglected because, under the condition of a weak probe field, the atomic operator $\hat{\sigma}_{23}(z,t)$, which is involved in the coupling field MSE, is negligible. By substituting Eq. (\ref{eq:(5)}) into Eqs. (\ref{eq:(4)}) and (\ref{eq:(6)}), we derive the equations of motion for $\hat{\sigma}_{12}(z, t)$ and $\hat{a}_p(z, t)$, which will be analyzed in the subsequent section.


\subsection{Dark-State Polariton} \label{subsec:DSP}

We employ the DSP construction introduced by Fleischhauer and Lukin \cite{DSP-1,DSP-2}. The dark polariton combines the probe-field amplitude and the collective ground-state coherence, while the orthogonal bright polariton describes the complementary excitation. Adiabatic variation of the coupling field changes the photonic and spin-wave components of the dark polariton, enabling storage and subsequent retrieval. This representation also provides a convenient starting point for determining the attenuation of the retrieved mode in the presence of finite ground-state coherence decay. For real $\Omega_c(t)=\Omega_c^*(t)$ and $g_p\in\mathbb{R}$, the dark-state polariton operator $\hat{\Psi}$ and the bright-state polariton operator $\hat{\Phi}$ are defined as
\begin{gather}
	\hat{\Psi}(z, t)={\rm cos}\theta(t)\hat{a}_p(z, t)-{\rm sin}\theta(t)\sqrt{N}\hat{\sigma}_{12}(z, t),\label{eq:(7)}\\
	\hat{\Phi}(z, t)={\rm sin}\theta(t)\hat{a}_p(z, t)+{\rm cos}\theta(t)\sqrt{N}\hat{\sigma}_{12}(z, t),\label{eq:(8)}
\end{gather}
where ${\rm tan}\theta(t) = \frac{2g_p\sqrt{N}}{\Omega_c(t)}$. The limits $\theta\simeq 0$ and $\theta\simeq \pi/2$ correspond to predominantly photonic and spin-wave excitations, respectively. The resulting change in the polariton composition during storage is central to the dependence of the attenuation factor on the coupling field. The equations of motion for $\hat{\sigma}_{12}(z,t)$ and $\hat{a}_p(z,t)$ are transformed into the new variables $\hat{\Psi}(z,t)$ and $\hat{\Phi}(z,t)$ using Eqs. (\ref{eq:(4)})--(\ref{eq:(6)}) as follows:
\begin{gather}
	\big(\frac{\partial}{\partial t}+c {\rm cos}^2\theta\frac{\partial}{\partial z}+\frac{\gamma_{21}'}{2}{\rm sin}^2\theta\big)\hat{\Psi}=\big(-\frac{\partial \theta}{\partial t}-\frac{c}{2}{\rm sin}2\theta\frac{\partial}{\partial z}+\frac{\gamma_{21}'}{4}{\rm sin}2\theta\big)\hat{\Phi}-\sqrt{N}{\rm sin}\theta\hat{F}_{12},\label{eq:(9)}\\
	\hat{\Phi}=\frac{{\rm sin}\theta}{2g_p^2 N}\big(\frac{\partial}{\partial t}+\frac{\gamma_{31}'}{2}\big)\Big[{\rm tan}\theta\big(2\frac{\partial}{\partial t}+\gamma_{21}'\big)\big({\rm sin}\theta\hat{\Psi}-{\rm cos}\theta\hat{\Phi}\big)+2\sqrt{N}{\rm tan}\theta\hat{F}_{12}\Big]+\frac{i{\rm sin}\theta}{g_p}\hat{F}_{13}.\label{eq:(10)}
\end{gather}
Here, we normalize time to the characteristic scale $T$ (the characteristic time) via $\widetilde{t}=t/T$ and expand Eqs. (\ref{eq:(9)}) and (\ref{eq:(10)}) in powers of $1/T$, or equivalently, the adiabaticity parameter $\epsilon \equiv (g_p\sqrt{N}T)^{-1}$. We first solve the homogeneous DSP evolution to determine the propagation and attenuation of the retrieved probe mode. The associated Langevin contribution is included when specifying the output quantum channel. To the lowest order in the adiabatic expansion \cite{DSP-2,DSP decoherence,Adiabatic limit-1,Adiabatic limit-2,Adiabatic limit-3}, the homogeneous equations become:
\begin{gather}
	\big(\frac{\partial}{\partial t}+c {\rm cos}^2\theta\frac{\partial}{\partial z}+\frac{\gamma_{21}'}{2}{\rm sin}^2\theta\big)\hat{\Psi}=\big(-\frac{c}{2}{\rm sin}2\theta\frac{\partial}{\partial z}+\frac{\gamma_{21}'}{4}{\rm sin}2\theta\big)\hat{\Phi},\label{eq:(11)}\\
	\hat{\Phi}=\frac{\gamma_{21}'\gamma_{31}'{\rm sin}^3\theta}{4g_p^2 N{\rm cos}\theta+\gamma_{21}'\gamma_{31}'{\rm sin}^2\theta{\rm cos}\theta}\hat{\Psi}.\label{eq:(12)}
\end{gather}
By substituting Eq. (\ref{eq:(12)}) into Eq. (\ref{eq:(11)}) and applying the spatial Fourier transform $\hat{f}(z, t)=\mathscr{F}^{-1}[\widetilde{f}(k, t)]=\int_{-\infty}^{\infty} \widetilde{f}(k, t)e^{-ikz}dk$, the equation of motion for the momentum-space DSP $\widetilde{\Psi}(k, t)$ is obtained as follows:
\begin{align}
	\frac{\partial\widetilde{\Psi}}{\partial t}=(ick{\rm cos}^2\theta-\frac{1}{2}\gamma_{21}'{\rm sin}^2\theta)\widetilde{\Psi}+(ick+\frac{\gamma_{21}'}{2})\frac{\gamma_{21}'\gamma_{31}'{\rm sin}^4\theta}{4g_p^2N+\gamma_{21}'\gamma_{31}'{\rm sin}^2\theta}\widetilde{\Psi}.\label{eq:(13)}
\end{align}
The momentum-space DSP can be explicitly solved from Eq. (\ref{eq:(13)}), and the position-space DSP is obtained by applying the inverse Fourier transform $\hat{\Psi}(z, t) = \mathscr{F}^{-1}[\widetilde{\Psi}(k, t)]$. The result is as follows:
\begin{gather}
	\hat{\Psi}(z, t)=e^{-\frac{\gamma_{21}'}{2}\int_0^t g_1(\tau)d\tau}\hat{\Psi}(z-c\int_0^t g_2(\tau)d\tau, 0),\label{eq:(14)}\\
	g_1(t)=\frac{4g_p^2 N}{\Omega_c^2(t)+4g_p^2 N+\gamma_{21}'\gamma_{31}'},\label{eq:(15)}\\
	g_2(t)=\frac{\Omega_c^2(t)+\gamma_{21}'\gamma_{31}'}{\Omega_c^2(t)+4g_p^2 N+\gamma_{21}'\gamma_{31}'}.\label{eq:(16)}
\end{gather}
For a strong coupling field, $\Omega_c^2 \gg 4g_p^2 N$, where $\theta \rightarrow 0$, the DSP exhibits a purely photonic character, $\hat{\Psi}(z, t) = \hat{a}_p(z, t)$, and propagates at the vacuum speed of light. In the opposite limit, with a weak coupling field, $\Omega_c^2 \ll 4g_p^2 N$, where $\theta \rightarrow \pi/2$, the DSP becomes spin-wave-like, $\hat{\Psi}(z, t) = -\sqrt{N}\hat{\sigma}_{12}(z, t)$, and its propagation speed approaches zero.

\FigTwo

Consider a time-dependent coupling field $\Omega_c(t)$ used for storing a probe pulse, as illustrated in Fig. \ref{fig:DSP propagation}. The Rabi frequency of the coupling field remains constant for $t \leq t_1$ and $t \geq t_2$, and the storage time is defined as $\Delta t_s = t_2 - t_1$. Assuming $\Omega_c^2(0) \gg 4g_p^2 N$, we have $\theta(t) \rightarrow 0$ in regions $\rm \uppercase\expandafter{\romannumeral 1}$ and $\rm \uppercase\expandafter{\romannumeral 3}$, where $\hat{\Psi}(z, t) \approx \hat{a}_p(z, t)$, indicating a photon-wave-dominated state. At $t = t_1$, as the coupling field strength decreases, the DSP gradually transforms from a photon-wave-dominated state into a spin-wave-dominated state. During the storage stage (region $\rm \uppercase\expandafter{\romannumeral 2}$), the coupling field is completely turned off, leading to $\theta(t) = \pi/2$, where $\hat{\Psi}(z, t) = -\sqrt{N}\hat{\sigma}_{12}(z, t)$. In the retrieval stage (region $\rm \uppercase\expandafter{\romannumeral 3}$), as the coupling field is turned back on, $\theta(t) \rightarrow 0$, and the retrieved DSP is converted back into a photon-wave-dominated state, $\hat{\Psi}(z, t) \approx \hat{a}_p(z, t)$. 

Equation~(\ref{eq:(14)}) determines the input-dependent component of the retrieved probe field. To evaluate its quantum state, we consider the quantum-limited loss regime in which the effective environmental mode is initially in vacuum and uncorrelated with the input, with no additional populated noise coupled into the detected output mode. In this regime the Langevin-driven contribution is represented by a normalized environmental mode, and the complete retrieved-field relation is
\begin{align}
	\hat{a}_p(z,t)=f(t)\hat{a}_p\left(z-c\int_0^t g_2(\tau)d\tau,0\right)+\sqrt{1-|f(t)|^2}\,\hat{v}_p(z,t),\label{eq:(17)}
\end{align}
where $f(t) \equiv e^{-\frac{\gamma_{21}'}{2}\int_0^t g_1(\tau)d\tau}$ is determined by the EIT dynamics and $\hat{v}_p$ is the independent environmental vacuum-mode operator. The coefficient $\sqrt{1-|f(t)|^2}$ preserves the canonical bosonic commutator of the output mode for $|f(t)|\leq 1$. The vacuum correlations, in particular $\langle\hat{v}_p^{\dagger}\hat{v}_p\rangle=0$, specify the quantum-limited reservoir considered here. Equation~(\ref{eq:(17)}) thus combines the EIT-derived attenuation with a well-defined quantum-limited noise model for the retrieved probe field.

In the general case where the coupling field $\Omega_{c}$ is not restricted to $\gg 4g_{p}^{2}N$, the probe pulse is converted into the DSP upon entering the EIT medium, undergoing significant spatial compression due to the reduced group velocity. Therefore, to avoid temporal deformation of the probe pulse upon retrieval, it is essential to ensure that the coupling field strength applied during the retrieval stage matches that applied at the input stage (region $\rm \uppercase\expandafter{\romannumeral 1}$) \cite{DSP-1, DSP-2}.


\section{Retrieved Quantum State} \label{sec:retrieved quantum state}

\subsection{Reduced Density Operator} \label{subsec:reduced density operator}

We derive the quantum state of the retrieved probe field using the reduced density operator theory \cite{Reduce Density Matrix-1}. In the Schr\"{o}dinger picture, the density operator of the combined system, consisting of the probe field and the reservoir, is expressed as:
\begin{align}
	&\rho_f=U\rho_i U^{\dagger}, \label{eq:(18)}
\end{align}
where $\rho_i = \rho_p(t=0) \otimes \rho_R(t=0)$ is the initial density operator, and $U(t)$ represents the time evolution operator of the combined system. For the quantum-limited regime considered here, the initial reservoir is in the vacuum state, $\rho_R(0)=|0_R\rangle\langle 0_R|$, and is uncorrelated with the probe. The unitary $U(t)$ represents a dilation of the resulting system--environment channel, which permits the reduced photonic state to be obtained by tracing over the environmental degrees of freedom. The full Heisenberg evolution of the probe ladder operator is written as
\begin{align}
	\widetilde{a}_{p}(k,t)
		=U^{\dagger}(t)[\widetilde{a}_p(k,0)\otimes I_R]U(t). \label{eq:(19)}
\end{align}
The density operator of the retrieved probe field in the Schr\"{o}dinger picture can be expanded in terms of the multimode number basis, and the corresponding density matrix elements are given as follows:
\begin{align}
	&\rho_{m_{k_i}\ldots m_{k_{i'}}, n_{k_j}\ldots n_{k_{j'}}}^p(t)\notag\\
	&=\prescript{}{p}{\langle}m_{k_i}\ldots m_{k_{i'}}|\rho_p(t)|n_{k_j}\ldots n_{k_{j'}}{\rangle}_p\notag\\
	&=\prescript{}{p}{\langle}m_{k_i}\ldots m_{k_{i'}}|tr_R(U\rho_iU^{\dagger})|n_{k_j}\ldots n_{k_{j'}}{\rangle}_p\notag\\
	&=tr_p\{|n_{k_j}\ldots n_{k_{j'}}{\rangle}_p\prescript{}{p}{\langle}m_{k_i}\ldots m_{k_{i'}}|tr_R(U\rho_iU^{\dagger})\}\notag\\
	&=tr_p\{tr_R[(|n_{k_j}\ldots n_{k_{j'}}{\rangle}_p\prescript{}{p}{\langle}m_{k_i}\ldots m_{k_{i'}}|\otimes I_R)U\rho_iU^{\dagger}]\}\notag\\
	&=tr\{U^{\dagger}(|n_{k_j}\ldots n_{k_{j'}}{\rangle}_p\prescript{}{p}{\langle}m_{k_i}\ldots m_{k_{i'}}|\otimes I_R)U\rho_i\}. \label{eq:(20)}
\end{align}
We define $\hat{\rho}_{m_{k_i}\ldots m_{k_{i'}}, n_{k_j}\ldots n_{k_{j'}}}^p(t)=U^{\dagger}(|n_{k_j}\ldots n_{k_{j'}}{\rangle}_p\prescript{}{p}{\langle}m_{k_i}\ldots m_{k_{i'}}|\otimes I_R)U$, which represents the Heisenberg picture operator for the density matrix element. For simplicity, in the subsequent discussion, we denote $\widetilde{a}_p(k_i, t) = \widetilde{a}_{k_i}(t)$. Utilizing the following expansion of the outer product of the multimode vacuum state \cite{Vacuum Outer Product Property}:
\begin{align}
	|0\rangle\langle 0|=\sum_{l=0}^{\infty}\frac{(-1)^l}{l!}(\frac{2\pi}{L})^l\int_{k_1}\ldots\int_{k_l}\widetilde{a}_{k_l}^{\dagger}(0)\ldots\widetilde{a}_{k_1}^{\dagger}(0)\widetilde{a}_{k_1}(0)\ldots\widetilde{a}_{k_l}(0)dk_1\ldots dk_l, \label{eq:(21)}
\end{align}
where $|0\rangle\langle 0| = |0_{k_i}\ldots 0_{k_{i'}}\rangle\langle 0_{k_i}\ldots 0_{k_{i'}}|$, we can express $|n_{k_j}\ldots n_{k_{j'}}{\rangle}_p \prescript{}{p}{\langle}m_{k_i}\ldots m_{k_{i'}}|$ as a sum involving the product of initial ladder operators. The detailed derivation of Eq. (\ref{eq:(21)}) is provided in Appendix \ref{sec:Appendix A}. Using this expansion, $\hat{\rho}_{m_{k_i}\ldots m_{k_{i'}}, n_{k_j}\ldots n_{k_{j'}}}^p(t)$ can be expressed as follows:
\begin{align}
	\hat{\rho}&_{m_{k_i}\ldots m_{k_{i'}}, n_{k_j}\ldots n_{k_{j'}}}^p(t)\notag\\
	=&U^{\dagger}(t)\bigg\{\sum_{l=0}^{\infty}\chi_{m_{k_i}\ldots m_{k_{i'}}, n_{k_j}\ldots n_{k_{j'}}, l}[\widetilde{a}_{k_j}^{\dagger}(0)]^{n_{k_j}}\ldots[\widetilde{a}_{k_{j'}}^{\dagger}(0)]^{n_{k_{j'}}}\Big[\int_{k_1}\ldots\int_{k_l}\widetilde{a}_{k_l}^{\dagger}(0)\ldots\widetilde{a}_{k_1}^{\dagger}(0)\notag\\
	&\times\widetilde{a}_{k_1}(0)\ldots\widetilde{a}_{k_l}(0)dk_1\ldots dk_l\Big][\widetilde{a}_{k_i}(0)]^{m_{k_i}}\ldots[\widetilde{a}_{k_{i'}}(0)]^{m_{k_{i'}}}\otimes I_R\bigg\}U(t)\notag\\
	=&\sum_{l=0}^{\infty}\chi_{m_{k_i}\ldots m_{k_{i'}}, n_{k_j}\ldots n_{k_{j'}}, l}[\widetilde{a}_{k_j}^{\dagger}(t)]^{n_{k_j}}\ldots[\widetilde{a}_{k_{j'}}^{\dagger}(t)]^{n_{k_{j'}}}\Big[\int_{k_1}\ldots\int_{k_l}\widetilde{a}_{k_l}^{\dagger}(t)\ldots\widetilde{a}_{k_1}^{\dagger}(t)\notag\\
	&\times\widetilde{a}_{k_1}(t)\ldots\widetilde{a}_{k_l}(t)dk_1\ldots dk_l\Big][\widetilde{a}_{k_i}(t)]^{m_{k_i}}\ldots[\widetilde{a}_{k_{i'}}(t)]^{m_{k_{i'}}}, \label{eq:(22)}
\end{align}
where $\chi_{m_{k_i}\ldots m_{k_{i'}}, n_{k_j}\ldots n_{k_{j'}}, l}\equiv\frac{(-1)^l}{l!}(\frac{2\pi}{L})^l\frac{1}{\sqrt{n_{k_j}!\ldots n_{k_{j'}}!m_{k_i}!\ldots m_{k_{i'}}!}}$, and the complete operator product is transformed under the full unitary evolution before the reservoir trace is taken. The momentum-space field operator in Eq. (\ref{eq:(22)}) can be obtained from Eq. (\ref{eq:(17)}) via a Fourier transform into the momentum domain. The phase shift arising from the spatial propagation of the probe field is neglected when comparing the input and retrieved quantum states. Thus, the momentum-space field operator is expressed as follows:
\begin{align}
	\widetilde{a}_k(t)
		=f(t)\widetilde{a}_k(0)
		+\sqrt{1-|f(t)|^2}\,\widetilde{v}_k(t),\label{eq:(23)}
\end{align}
where $\widetilde{v}_k(t)$ represents the corresponding vacuum-environment mode in momentum space. We treat the retrieved modes as independent within the present approximation and neglect additional mode coupling and distortion beyond those captured by the DSP dynamics. Equation~(\ref{eq:(23)}) combines the EIT-derived attenuation of the retrieved probe mode with the environmental contribution required to preserve the canonical bosonic commutation relation. Under the specified quantum-limited vacuum-reservoir conditions, the resulting input--output relation takes the form of a pure-loss channel. The storage efficiency is defined as $\eta\equiv\frac{\langle\hat{n}(t)\rangle}{\langle\hat{n}(0)\rangle}$. With the reservoir initially in the vacuum state, the noise contribution vanishes in normally ordered photon-number expectation values, yielding the intrinsic retrieved-mode efficiency
\begin{align}
	\eta=\frac{\int_{-\infty}^{\infty}\langle\widetilde{a}_k^{\dagger}(t)\widetilde{a}_k(t)\rangle dk}{\int_{-\infty}^{\infty}\langle\widetilde{a}_{k'}^{\dagger}(0)\widetilde{a}_{k'}(0)\rangle dk'}=|f(t)|^2.\label{eq:(24)}
\end{align}
Equations~(\ref{eq:(23)}) and~(\ref{eq:(24)}) show that, under the assumptions adopted here, the retrieved photonic mode is described by an effective pure-loss channel. The loss parameter is not introduced phenomenologically, but is determined by the DSP evolution and by ground-state decoherence through the attenuation factor $f(t)$, while the corresponding pure-loss form is adopted within the specified quantum-limited vacuum-reservoir model. This connection enables the EIT-dependent attenuation to be incorporated into the density-operator calculations for single-photon and multimode input states below.

Note that the integration range in $f(t)$, as defined in Eq.~(\ref{eq:(17)}), can be approximated in the region $\rm \uppercase\expandafter{\romannumeral 2}$ where the coupling field is turned off. In this case, the coefficient $f(t)\approx e^{-\frac{\gamma_{21}'}{2}\frac{4g_p^2N}{4g_p^2N+\gamma_{21}'\gamma_{31}'}\Delta t_s}$, which is originally a function of time, becomes a constant when the storage time $\Delta t_s$ is fixed. Since the operator products in Eq.~(\ref{eq:(22)}) are normally ordered, all terms containing environmental operators vanish upon tracing over the vacuum reservoir. The remaining input-field contribution gives
\begin{align}
	\rho&_{m_{k_i}\ldots m_{k_{i'}}, n_{k_j}\ldots n_{k_{j'}}}^p(t)\notag\\
	=&\sum_{l=0}^{\infty}\chi_{m_{k_i}\ldots m_{k_{i'}}, n_{k_j}\ldots n_{k_{j'}}, l}[f^*(t)]^{n_{k_j}+\ldots+n_{k_{j'}}+l}[f(t)]^{m_{k_i}+\ldots+m_{k_{i'}}+l}\notag\\
	&\times{\rm tr}_p\bigg\{[\widetilde{a}_{k_j}^{\dagger}(0)]^{n_{k_j}}\ldots[\widetilde{a}_{k_{j'}}^{\dagger}(0)]^{n_{k_{j'}}}\Big[\int_{k_1}\ldots\int_{k_l}\widetilde{a}_{k_l}^{\dagger}(0)\ldots\widetilde{a}_{k_1}^{\dagger}(0)\notag\\
	&\times\widetilde{a}_{k_1}(0)\ldots\widetilde{a}_{k_l}(0)dk_1\ldots dk_l\Big][\widetilde{a}_{k_i}(0)]^{m_{k_i}}\ldots[\widetilde{a}_{k_{i'}}(0)]^{m_{k_{i'}}}\rho_p(0)\bigg\}.\label{eq:(25)}
\end{align}


\subsection{Multimode Fock State} \label{subsec:multimode Fock state}

We first consider the case where the input probe field is in a multimode single-photon Fock state, $|1; \phi_p'(k, 0)\rangle = \sqrt{\frac{2\pi}{L}} \int_{-\infty}^{\infty} \phi_p'(k, 0) |1_{k, 0}\rangle dk$, and determine the density operator of the retrieved probe field after it has been stored in the $\Lambda$-type atomic medium. Details regarding the multimode single-photon Fock state are provided in Appendix \ref{sec:Appendix B}. Using Eq. (\ref{eq:(25)}), the density matrix elements of the retrieved probe field are derived as follows:
\begin{align}
	\rho_{m_{k_i}\ldots m_{k_{i'}}, n_{k_j}\ldots n_{k_{j'}}}^p(t)=&\delta_{m_{k_i}+\ldots+m_{k_{i'}}, 0}\delta_{n_{k_j}+\ldots+n_{k_{j'}}, 0}(1-\eta)+\delta_{m_{k_i}+\ldots+m_{k_{i'}}, 1}\delta_{n_{k_j}+\ldots+n_{k_{j'}}, 1}\eta\notag\\
	&\times\frac{L}{2\pi}{\phi_p'}^*(k_q, 0)|_{n_{k_q}=1}\phi_p'(k_p, 0)|_{m_{k_p}=1}.\label{eq:(26)}
\end{align}
Since the density matrix elements of the input probe field are
\begin{align}
	\rho_{m_{k_i}\ldots m_{k_{i'}}, n_{k_j}\ldots n_{k_{j'}}}^p(0)=\delta_{m_{k_i}+\ldots+m_{k_{i'}}, 1}\delta_{n_{k_j}+\ldots+n_{k_{j'}}, 1}\frac{L}{2\pi}{\phi_p'}^*(k_q, 0)|_{n_{k_q}=1}\phi_p'(k_p, 0)|_{m_{k_p}=1},\label{eq:(27)}
\end{align}
the density operator for the retrieved probe field is expressed as follows:
\begin{align}
	\rho_p=(1-\eta)|0\rangle\langle 0|+\eta|1; \phi_p'(k, 0)\rangle\langle 1; \phi_p'(k, 0)|,\label{eq:(28)}
\end{align}
which represents a mixed state comprising the multimode vacuum state and the original single-photon Fock state $|1; \phi_p'(k, 0)\rangle$. The fidelity between the input and retrieved quantum states (as defined in \cite{Fidelity}), also referred to as the storage fidelity, is given by:
\begin{align}
	F=\sqrt{\langle 1; \phi_p'(k, 0)|\rho_p(t)|1; \phi_p'(k, 0)\rangle}=\sqrt{\eta}.\label{eq:(29)}
\end{align}
For the fully resonant EIT quantum memory system (i.e., $\Delta_p = \delta = 0$), the storage fidelity depends only on the ground-state decoherence rate $\gamma_{21}$ and the storage time $\Delta t_s$, as the storage efficiency is given by $\eta \approx e^{-\gamma_{21}\Delta t_s}$ under the condition of a strong coupling field and a weak ground state decoherence rate ($4g_p^2N\gg\gamma_{21}\gamma_{31}$). The fidelity defined in Eq.~(\ref{eq:(29)}) is an unconditional fidelity evaluated for the retrieved photonic state and includes the vacuum component generated by storage loss. It therefore differs from post-selected fidelities commonly reported in experiments, where loss primarily reduces the success probability and the detected non-vacuum subspace may retain high state fidelity~\cite{Maximally Path-Entangled Single-Photon Quantum Memory Atomic Ensemble,Exp Fidelity-1,Exp Fidelity-2,Exp Fidelity-3,Exp Fidelity-4}. From a quantum mechanical perspective, photon loss transforms a single-photon state into a mixture of the single-photon and vacuum states, thereby reducing the unconditional fidelity. In addition, our fidelity omits the phase fluctuation effect that can occur during photon propagation through optical fibers (as considered in, e.g., Ref.~\cite{Exp Fidelity-5}) and focuses on the internal decoherence behavior of the quantum memory.

\FigThree

We now consider a practical quantum memory system based on $\Lambda$-type EIT, where the energy levels of $^{87}\text{Rb}$ are chosen as $|1\rangle = |5S_{1/2}, F=2, m_F=2\rangle$, $|2\rangle = |5S_{1/2}, F=1, m_F=0\rangle$, and $|3\rangle = |5P_{3/2}, F=1, m_F=1\rangle$. For this system, both the probe and coupling fields lie near the $^{87}\text{Rb}$ $D_2$ transition at 780 nm, with their frequencies separated primarily by the ground-state hyperfine splitting. The probe field drives the $\sigma^{-}$ transition $|1\rangle\leftrightarrow|3\rangle$, while the coupling field drives the $\sigma^{+}$ transition $|2\rangle\leftrightarrow|3\rangle$. In this resonant EIT system, the theoretically calculated fidelity between the input and retrieved probe states is shown in Fig.~\ref{fig:multimode Fock state fidelity}. We further discuss the storage of a single-rail qubit, where the two-dimensional Hilbert space is spanned by the vacuum state $|0\rangle$ and the single-photon Fock state $|1; \phi_p'(k, 0)\rangle$ within a spatial mode of the electromagnetic field \cite{Photon Number Qubit Teleportation-1, Photon Number Qubit Teleportation-2}. Suppose the probe field is prepared in an arbitrary one-qubit state using single-rail encoding. The density operator, represented in the logical basis ($|0\rangle \equiv |0\rangle$ and $|1\rangle \equiv |1; \phi_p'(k, 0)\rangle$), is given as follows:
\begin{align}
	\rho_p(0)=
	\begin{bmatrix}
		\rho_{00} & \rho_{01}\\
		\rho_{10} & \rho_{11}
	\end{bmatrix}.\label{eq:(30)}
\end{align}
While Eq. (\ref{eq:(30)}) represents the input probe field in the two-dimensional Hilbert space spanned by the vacuum and single-photon states, this specific setting corresponds to the single-rail qubit case discussed above. The general form of our theoretical framework, particularly Eq. (\ref{eq:(25)}), is not restricted to this case and can also accommodate other quantum input states, such as coherent states or squeezed states. For these more general inputs, the output states and the evolution of the density matrix can still be computed within the same formalism, although analytic expressions may no longer be available and numerical methods may be required.

Using Eq. (\ref{eq:(25)}), the exact form of the density matrix elements of the retrieved probe field is given as follows:
\begin{align}
	\rho&_{m_{k_i}\ldots m_{k_{i'}}, n_{k_j}\ldots n_{k_{j'}}}^p(t)\notag\\
	=&\delta_{m_{k_i}+\ldots+m_{k_{i'}}, 0}\delta_{n_{k_j}+\ldots+n_{k_{j'}}, 0}[\rho_{00}+(1-|f(t)|^2)\rho_{11}]\notag\\
	&+\delta_{m_{k_i}+\ldots+m_{k_{i'}}, 0}\delta_{n_{k_j}+\ldots+n_{k_{j'}}, 1}f^*(t)\rho_{01}\sqrt{\frac{L}{2\pi}}{\phi_p'}^*(k_q, 0)|_{n_{k_q}=1}\notag\\
	&+\delta_{m_{k_i}+\ldots+m_{k_{i'}}, 1}\delta_{n_{k_j}+\ldots+n_{k_{j'}}, 0}f(t)\rho_{10}\sqrt{\frac{L}{2\pi}}\phi_p'(k_p, 0)|_{m_{k_p}=1}\notag\\
	&+\delta_{m_{k_i}+\ldots+m_{k_{i'}}, 1}\delta_{n_{k_j}+\ldots+n_{k_{j'}}, 1}|f(t)|^2\rho_{11}\frac{L}{2\pi}{\phi_p'}^*(k_q, 0)|_{n_{k_q}=1}\phi_p'(k_p, 0)|_{m_{k_p}=1}.\label{eq:(31)}
\end{align}
The density operator of the retrieved probe field, represented in the logical basis, is given as follows:
\begin{align}
	\rho_p =
	\begin{bmatrix}
		\rho_{00}+(1-|f(t)|^2)\rho_{11} & f^*(t)\rho_{01}\\
		f(t)\rho_{10} & |f(t)|^2\rho_{11}
	\end{bmatrix}.\label{eq:(32)}
\end{align}
From Eq.~(\ref{eq:(32)}), the validity of the present model can be checked by examining known limits. In the absence of decoherence, $f(t) = 1$ and the retrieved probe density matrix exactly reproduces the input state. In the strong-decoherence limit, $f(t) \to 0$ and the retrieval fidelity approaches that of the vacuum projection, consistent with energy-loss expectations. These limiting behaviors are physically intuitive and consistent with the phenomenology of decoherence in EIT-based quantum memory, although direct theoretical treatments for the impact of ground-state decoherence on the storage process remain sparse. Equation~(\ref{eq:(32)}) represents the ground-state-decoherence-limited pure-loss limit of the EIT memory model. In this limit, the retained effect of ground-state decoherence on the retrieved photonic state appears as vacuum admixture induced by loss. Additional noise or mode-imperfection mechanisms are not included in the present model and would require an extended treatment of the input-output relation and the retrieved density operator. Note that the total population in the photonic subspace $\{|0\rangle, |1,\phi_p'(k,0)\rangle\}$ remains constant because the decohered portion of the spin wave is mapped to the vacuum component of the field, rather than indicating energy conservation in the full atom-field system.


\section{Entanglement and Clauser–Horne Tests} \label{sec:entanglement retention}

\subsection{Multiple Qubits} \label{subsec:multiple qubits}

In this section, we extend the quantum memory system to store an arbitrary number of qubits. We first consider a quantum memory system comprising $N$ spatially separated atomic ensembles, where each ensemble functions as a $\Lambda$-type EIT memory and corresponds to the storage of a spatial mode of the probe field. The input probe field of the entire system is treated as a whole, and its density operator $\rho_p(0)$, including all spatial modes of the probe field, is taken as the input state. Here, the overall input probe field is allowed to be an arbitrary multi-photon mixed state of spatial modes. Using the reduced density operator theory, we derive the density matrix elements of the retrieved probe field as follows:
\begin{align}
	\rho&_{m_{k_i, 1}\ldots m_{k_{i'}, 1}\ldots m_{k_i, N}\ldots m_{k_{i'}, N}, n_{k_j, 1}\ldots n_{k_{j'}, 1}\ldots n_{k_j, N}\ldots n_{k_{j'}, N}}^p(t)\notag\\
	=&{\rm tr}_p\bigg\{\prod_{r=1}^N\sum_{l_r=0}^{\infty}\chi_{m_{k_i, r}\ldots m_{k_{i'}, r}n_{k_j, r}\ldots n_{k_{j'}, r}l_r}[f_r^*(t)]^{n_r+l_r}[f_r(t)]^{m_r+l_r}[\widetilde{a}_{k_j, r}^{\dagger}(0)]^{n_{k_j, r}}\ldots\notag\\
	&\times[\widetilde{a}_{k_{j'}, r}^{\dagger}(0)]^{n_{k_{j'}, r}}\int_{k_{1, r}}\ldots\int_{k_{l_r, r}}\widetilde{a}_{k_{l_r, r}, r}^{\dagger}(0)\ldots\widetilde{a}_{k_{1, r}, r}^{\dagger}(0)\widetilde{a}_{k_{1, r}, r}(0)\ldots\widetilde{a}_{k_{l_r, r}, r}(0)\notag\\
	&\times dk_{1, r}\ldots dk_{l_r, r}[\widetilde{a}_{k_i, r}(0)]^{m_{k_i, r}}\ldots[\widetilde{a}_{k_{i'}, r}(0)]^{m_{k_{i'}, r}}\rho_p(0)\bigg\}\notag\\
	\equiv&{\rm tr}_p\{\prod_{r=1}^N\hat{O}_{m_{k_i, r}\ldots m_{k_{i'}, r}n_{k_j, r}\ldots n_{k_{j'}, r}r}\rho_p(0)\},\label{eq:(33)}
\end{align}
where $m_r \equiv m_{k_i, r} + \ldots + m_{k_{i'}, r}$ and $n_r \equiv n_{k_j, r} + \ldots + n_{k_{j'}, r}$ represent the photon numbers for the states $|m_{k_i} \ldots m_{k_{i'}}\rangle_r$ and $|n_{k_j} \ldots n_{k_{j'}}\rangle_r$, respectively. The derivation here follows the same procedure as in Sec.~\ref{sec:retrieved quantum state} and makes no assumption on the specific input state, thereby yielding a general expression valid for arbitrary multi-photon, multi-mode photonic states. Thus, Eq.~(\ref{eq:(33)}) extends the single-memory reduced-density-operator treatment to $N$ spatially separated EIT memories, with each spatial mode characterized by its own ground-state-decoherence-induced attenuation factor.

Consider an $N$-qubit input state $\rho_p(0)$, physically implemented using $N$ spatial modes of the electromagnetic field, where each qubit is encoded in the photon-number DOF of its corresponding spatial mode. The $i$-th qubit is sent into the $i$-th atomic ensemble and converted into a spin-wave during the storage stage. The density operator of the input probe field can be expanded using the logical basis representation as follows:
\begin{align}
	\rho_p(0)=\sum_{i_1=0}^1\ldots\sum_{i_N=0}^1\sum_{j_1=0}^1\ldots\sum_{j_N=0}^1\rho_{i_1\ldots i_N, j_1\ldots j_N}^p(0)|i_1\ldots i_N\rangle\langle j_1\ldots j_N|,\label{eq:(34)}
\end{align}
where the logical basis of the $i$-th field mode is defined as $|0_i\rangle = |0\rangle_i$ and $|1_i\rangle = |1; \phi_{p, i}'(k, 0)\rangle_i$. In the discrete-variable case considered here, Eq.~(\ref{eq:(34)}) restricts each spatial mode to carry either zero or one photon, corresponding to a general discrete-variable $N$-qubit state. Using Eq.~(\ref{eq:(33)}) and leveraging the separability of operators for distinct ensembles, we obtain:
\begin{align}
	\rho&_{m_{k_i, 1}\ldots m_{k_{i'}, 1}\ldots m_{k_i, N}\ldots m_{k_{i'}, N}, n_{k_j, 1}\ldots n_{k_{j'}, 1}\ldots n_{k_j, N}\ldots n_{k_{j'}, N}}^p(t)\notag\\
	=&\sum_{i_1=0}^1\ldots\sum_{i_N=0}^1\sum_{j_1=0}^1\ldots\sum_{j_N=0}^1\rho_{i_1\ldots i_N, j_1\ldots j_N}^p(0)\prod_{r=1}^N\langle j_r|\hat{O}_{m_{k_i, r}\ldots m_{k_{i'}, r}n_{k_j, r}\ldots n_{k_{j'}, r}r}|i_r\rangle\notag\\
	=&\sum_{i_1=0}^1\ldots\sum_{i_N=0}^1\sum_{j_1=0}^1\ldots\sum_{j_N=0}^1\rho_{i_1\ldots i_N, j_1\ldots j_N}^p(0)\prod_{r=1}^N\Big\{\delta_{m_rn_r, 00}\big[\delta_{i_rj_r, 00}+\delta_{i_rj_r, 11}(1-|f_r(t)|^2)\big]\notag\\
	&+\delta_{m_rn_r, 01}\delta_{i_rj_r, 01}\sqrt{\frac{L}{2\pi}}{\phi_r'}^*(k_u, 0)f_r^*(t)+\delta_{m_rn_r, 10}\delta_{i_rj_r, 10}\sqrt{\frac{L}{2\pi}}\phi_r'(k_v, 0)f_r(t)\notag\\
	&+\delta_{m_rn_r, 11}\delta_{i_rj_r, 11}\frac{L}{2\pi}{\phi_r'}^*(k_u, 0)\phi_r'(k_v, 0)|f_r(t)|^2\Big\}|_{n_{k_u, r}=1, m_{k_v, r}=1}.\label{eq:(35)}
\end{align}
Here, for simplicity, we denote $\phi_{p, r}'(k, 0)$ as $\phi_r'(k, 0)$. Equation~(\ref{eq:(35)}) gives the explicit transformation of a general discrete-variable $N$-qubit photonic state when each spatial mode is stored in and retrieved from an independent EIT memory. The result is the multi-spatial-mode extension of the single-memory density-operator map, with each factor $f_r(t)$ determined by the corresponding EIT storage process and ground-state decoherence. If $f_r(t) = 1$ (i.e., $\eta_r = 1$) for all $r \in \{1, \ldots, N\}$—corresponding to $\gamma_{21}' = \gamma_{21} - 2i\delta = 0$ and thus no ground-state decoherence $\gamma_{21}$ and zero two-photon detuning $\delta$—then $\rho_p(t)=\rho_p(0)$, meaning the retrieved density operator is identical to that of the input field. This is the expected noiseless-memory limit, in which the multi-memory formalism reduces to perfect state recovery when ground-state decoherence and two-photon detuning are absent.

For the storage of a path-encoded $N$-qubit state, following a similar discussion as in \cite{Diamond-type QFC}, the result parallels the single-rail case. The retrieved density matrix elements retain the form in Eq.~(\ref{eq:(35)}), indicating that the same density-operator transformation applies to path encoding, with perfect state recovery reached in the noiseless-memory limit. For polarization-encoded $N$-qubit states, a polarization-to-path conversion via polarization beam splitters allows the same independent-memory description to be applied.

Our framework analyzes a $\Lambda$-type ensemble EIT memory and, in the discrete-variable specialization considered here, already treats multi-mode photonic states and yields the retrieved density operator. This differs from single-ensemble quantum multiplexing, which requires multiple independently addressable, mutually orthogonal channels (e.g., temporal bins, frequency bins tied to distinct Raman resonances or $\Lambda$ transitions, polarization double-$\Lambda$, or spatial/OAM modes) with low crosstalk and stable phase control. Related theoretical studies have explored quantum multiplexing, entanglement distribution, and multiqubit entanglement generation in quantum networks \cite{Quantum multiplexing-1,Quantum multiplexing-2,Quantum multiplexing-3,Quantum multiplexing-4}. Our framework could in principle be extended to incorporate single-ensemble multiplexing once such channels are engineered. Modeling multiplexed or interacting memory channels would require incorporating Hamiltonian couplings, mode overlap, inter-channel crosstalk, and cross-decoherence among collective spin-wave modes, which are outside the present independent-memory model.

\FigFour


\subsection{Single-Photon Entanglement} \label{subsec:single-photon entanglement}

We provide a detailed analysis on the storage of the following maximally path-entangled single-photon state \cite{Maximally Path-Entangled Single-Photon Quantum Memory Atomic Ensemble}:
\begin{align}
	|\phi_p(0)\rangle=\frac{1}{\sqrt{2}}(|1_A0_B\rangle-|0_A1_B\rangle),\label{eq:(36)}
\end{align}
where the logical basis is defined as the single-mode vacuum state $|0\rangle_{k_0}$ and the single-mode Fock state $|1\rangle_{k_0}$. Note that the single mode in momentum-space $k_0$ is equivalent to the single mode in frequency-space $\omega_0 = ck_0$. The density operator of the output probe field can be obtained using the single-mode version of Eq. (\ref{eq:(35)}). As shown in Fig. \ref{fig:CH inequality}(a), we add a phase shifter $\widetilde{P}(\phi_{A (B)})$ to each port to eliminate the phase of the output probe field. Specifically, $\widetilde{a}_{k_0, A (B)}'(t) = e^{-i\phi_{A (B)}}\widetilde{a}_{k_0, A (B)}(t) = |f_{A (B)}(t)|\widetilde{a}_{k_0, A (B)}(0)+\sqrt{1-|f_{A (B)}(t)|^2}\,\widetilde{v}_{k_0, A (B)}'(t)$. Here, $\widetilde{v}_{k_0, A (B)}'(t)=e^{-i\phi_{A (B)}}\widetilde{v}_{k_0, A (B)}(t)$ denotes the phase-adjusted environmental vacuum-mode operator, with independent vacuum modes for the two spatial paths. The density operator of the probe field after the phase shifter is then derived as follows:
\begin{align}
	\rho_p=&\frac{1}{2}\eta_A|1_A0_B\rangle\langle 1_A0_B|-\frac{1}{2}\sqrt{\eta_A\eta_B}|1_A0_B\rangle\langle 0_A1_B|-\frac{1}{2}\sqrt{\eta_A\eta_B}|0_A1_B\rangle\langle 1_A0_B|\notag\\
	&+\frac{1}{2}\eta_B|0_A1_B\rangle\langle 0_A1_B|+(1-\frac{1}{2}\eta_A-\frac{1}{2}\eta_B)|0_A0_B\rangle\langle 0_A0_B|.\label{eq:(37)}
\end{align}
Equation~(\ref{eq:(37)}) is the unconditional retrieved density operator for the two spatial modes and explicitly includes the vacuum component generated by storage loss. Thus, storage loss reduces the weight of the single-photon entangled sector in the unconditional state, but the presence of vacuum admixture alone does not imply that the surviving non-vacuum component is separable. For the symmetric pure-loss case, $\eta_A=\eta_B=\eta$, Eq.~(\ref{eq:(37)}) reduces to $\rho_p=(1-\eta)|0_A0_B\rangle\langle0_A0_B|+\eta|\phi_p(0)\rangle\langle\phi_p(0)|$, which remains entangled for any $\eta>0$. For $\eta<1$, however, the unconditional output is no longer the original pure Bell state but a statistical mixture of the vacuum and the Bell component. For the symmetric case, conditioning on successful retrieval selects the single-photon sector, whose normalized state is $|\phi_p(0)\rangle$.

Using the root-fidelity convention, $F=\sqrt{\langle\phi_p(0)|\rho_p|\phi_p(0)\rangle}$, the fidelity between the retrieved probe field and the input probe field is given as follows:
\begin{align}
	F=\frac{1}{2}(\sqrt{\eta_A}+\sqrt{\eta_B}).\label{eq:(38)}
\end{align}
The fidelity is depicted in Fig. \ref{fig:CH inequality}(b) as a function of $\eta_A$ and $\eta_B$. This fidelity is evaluated for the unconditional retrieved photonic state, including the vacuum component generated by loss. It therefore quantifies how storage loss reduces the overlap between the input Bell state and the full retrieved density operator, rather than the fidelity of a post-selected state conditioned on photon detection.


\subsection{Clauser–Horne Inequality} \label{Clauser-Horne Tests}

To investigate the nonlocal behavior between the two spatial modes of the retrieved probe field, we consider the unbalanced homodyne detection scheme \cite{Single-photon nonlocality-1, Single-photon nonlocality-2}, as depicted in Fig. \ref{fig:CH inequality}(a). Throughout this section we assume an ideal, lossless beam splitter (unitary transformation) and ideal non-number-resolving on/off detectors with unit efficiency and negligible dark counts, since our focus is on the intrinsic decoherence of the EIT memory rather than imperfections in the measurement apparatus. In the limit of a highly transmissive beam splitter ($T \rightarrow 1$) and assuming the local oscillator state $|\gamma\rangle_{k_0}$ is extremely strong ($|\gamma| \rightarrow \infty$), the beam splitter can be effectively described by the displacement operator $\hat{D}(\sqrt{1-T}\gamma) = \hat{D}(\alpha)$ \cite{Detailed on the POVM}, and we follow the positive operator-valued measure (POVM) conventions therein for displaced on/off detection, expressing the elements in normal-ordered form. It is convenient to quantify the displacement by $|\alpha|^{2}$. In the CH settings we use $\alpha\in\{0,\alpha'\}$ for mode $A$ and $\beta\in\{0,\beta'\}$ with $\beta'=-\alpha'$ for mode $B$ (implemented by setting the relative local oscillator phase at mode $B$ to $\pi$), and we define $J \equiv |\alpha'|^{2}$ with $\alpha'=\sqrt{1-T}\,\gamma$. The displacement operator $\hat{D}(\alpha)$ effectively shifts the retrieved probe mode in phase space by $\alpha$, enabling the measurement of quantum interference between the local oscillator and the probe field. The probabilities defined below include both click and no-click events explicitly. Therefore, the CH test considered here is applied to the unconditional retrieved state, including the vacuum component generated by storage loss, without post-selection on photon-detection events. The corresponding local POVM is given as follows:
\begin{gather}
	\hat{Q}(\alpha)=\hat{D}(\alpha)|0\rangle\langle 0|\hat{D}^{\dagger}(\alpha),\label{eq:(39)}\\
	\hat{P}(\alpha)=\hat{D}(\alpha)\sum_{n=1}^{\infty}|n\rangle\langle n|\hat{D}^{\dagger}(\alpha).\label{eq:(40)}
\end{gather}
Here, $\hat{Q}(\alpha)+\hat{P}(\alpha)=\hat{I}$, ensuring the completeness of the measurement operators. The joint probability of a no-click event occurring simultaneously at both photon detectors is given by:
\begin{align}
	Q_{ab}(\alpha, \beta)={\rm tr}[\rho_p(t)\hat{Q}_a(\alpha)\otimes\hat{Q}_b(\beta)],\label{eq:(41)}
\end{align}
The probabilities of no-click events at individual detectors are given by:
\begin{gather}
	Q_{a}(\alpha)={\rm tr}[\rho_p(t)\hat{Q}_a(\alpha)\otimes\hat{I}_b],\label{eq:(42)}\\
	Q_{b}(\beta)={\rm tr}[\rho_p(t)\hat{I}_a\otimes\hat{Q}_b(\beta)].\label{eq:(43)}
\end{gather}
The probabilities for single-detector clicks and coincidence clicks are $P_a(\alpha)=1-Q_a(\alpha)$, $P_b(\beta)=1-Q_b(\beta)$, and $P_{ab}(\alpha, \beta)=1-Q_a(\alpha)-Q_b(\beta)+Q_{ab}(\alpha, \beta)$. In the CH Bell test \cite{Single-photon nonlocality-1, Single-photon nonlocality-2}, measurements are performed under two different settings of coherent displacements for each mode ($A$ or $B$). A detector click is assigned a value of 1, while a no-click event is assigned a value of 0. With the above configuration $(\alpha,\beta)\in\{(0,0),(0,-\alpha'),(\alpha',0),(\alpha',-\alpha')\}$ and $J=|\alpha'|^{2}$, the CH combination is constructed as follows:
\begin{align}
	CH=&P_{ab}(0, 0)+P_{ab}(0, -\alpha')+P_{ab}(\alpha', 0)-P_{ab}(\alpha', -\alpha')-P_a(0)-P_b(0),\label{eq:(44)}
\end{align}
which satisfies the CH inequality, $-1 \leq CH \leq 0$, as predicted by local hidden variable theories \cite{EPR Paradox, Bell’s Theorem and Locally Mediated Models}. The CH combination for the two spatial modes of the retrieved probe field is obtained by substituting Eq. (\ref{eq:(37)}) into Eq. (\ref{eq:(44)}), resulting in the following expression:
\begin{align}
	CH=&-1+e^{-J}[2-\eta_A-\eta_B+\frac{J}{2}(\eta_A+\eta_B)]-e^{-2J}[1-\frac{1}{2}(\eta_A+\eta_B)+\frac{J}{2}(\sqrt{\eta_A}+\sqrt{\eta_B})^2].\label{eq:(45)}
\end{align}
Using Eq. (\ref{eq:(45)}), the minimum value of $CH$ for each set of $\eta_A$ and $\eta_B$ can be determined by tuning the value of the coherent displacement $J$. The results are shown in Fig. \ref{fig:CH inequality}(c). Storage processes with $CH_{\text{min}} < -1$ signify a violation of the CH inequality for specific values of $J$, demonstrating unconditional CH nonlocality \cite{Bell Nonlocality} between the two field modes $A$ and $B$ under the specified displaced on/off detection protocol without post-selection. For the symmetric case with $\eta_A=\eta_B=\eta$, the model gives a protocol-dependent intrinsic storage-efficiency benchmark of $\eta=89.7\%$ for violating the chosen unconditional CH inequality. This value depends on the specific CH inequality, displaced on/off detection scheme, ideal detector model, and absence of post-selection, and should not be interpreted as a universal threshold for EIT memories or for entanglement preservation. Below this benchmark, the unconditional retrieved state no longer violates the chosen CH inequality within this protocol. This does not imply that the surviving non-vacuum component is separable, nor does it exclude conditional or post-selected demonstrations of quantum correlations. In the pure-loss limit considered here, storage loss primarily reduces the success probability and the weight of the single-photon sector in the unconditional state.

The CH analysis therefore addresses a different question from conditional or heralded retrieval. It examines whether the complete retrieved state, including the vacuum component generated by storage loss, still violates the specified CH inequality under displaced on/off detection without conditioning on successful retrieval. The contrast between entanglement of the complete retrieved state persisting for any $\eta > 0$ and unconditional CH violation requiring $\eta > 89.7\%$ in the symmetric quantum-limited pure-loss case shows that the specified unconditional CH test imposes a more stringent requirement on storage efficiency than entanglement survival. This intrinsic benchmark isolates the EIT-memory loss considered in the present ground-state-decoherence-limited model. The quoted 89.7\% value should therefore not be interpreted as an end-to-end experimental efficiency requirement, for which additional experimental losses and imperfections would also need to be considered.


\section{Conclusion} \label{sec:conclusion}

In summary, we develop a theoretical framework for EIT quantum memories that incorporates ground-state-decoherence-induced loss by combining the DSP formalism with reduced density operator theory. Within the weak-probe and adiabatic-storage approximations, and under the specified quantum-limited vacuum-reservoir conditions, the resulting input-output relation can be interpreted as an effective pure-loss channel parameterized by EIT-derived attenuation for the retrieved photonic state. The derived density operator shows that ground-state decoherence leads to loss of the stored probe photon, transforming an initially pure Bell state into a mixed state with a vacuum component. The complete retrieved mixed state nevertheless remains entangled when both path-mode storage efficiencies are nonzero. Our analysis identifies a protocol-dependent storage-efficiency benchmark of 89.7\% for violating the chosen unconditional CH inequality with displaced on/off detection and no post-selection. Above this benchmark, the unconditional retrieved photonic state violates the CH inequality within the specified measurement protocol, whereas below it the chosen unconditional CH violation is no longer obtained. This benchmark is protocol-dependent rather than a universal threshold for other Bell tests, measurement protocols, or entanglement witnesses. Extending the model to multiple EIT memories further yields the retrieved $N$-qubit photonic density operator under independent ground-state-decoherence-induced loss, with the ideal noiseless-memory limit recovering the input state.

Beyond these quantitative results, this work provides a theoretical description of how entanglement retention and unconditional CH nonlocality evolve in EIT-based quantum memories subject to ground-state-decoherence-induced loss. The framework connects the EIT atom--field dynamics and ground-state decoherence to the properties of the retrieved photonic quantum state, thereby clarifying how the effective loss parameter, the retrieved density operator, and the unconditional CH benchmark are related. In this sense, the attenuation factor is determined by the EIT/DSP equations rather than introduced as an independent phenomenological loss parameter, whereas the quantum-limited pure-loss form is specified by the effective reservoir model. The present model describes the ground-state-decoherence-limited pure-loss regime with effective Markovian relaxation and does not include additional noise or mode-imperfection mechanisms. Different microscopic relaxation mechanisms may involve different Langevin noise correlations, which can be considered in future extensions. These results extend the DSP description by explicitly relating it to the retrieved photonic density operator and provide a useful theoretical basis for analyzing unconditional retrieved-state fidelity, entanglement retention, and protocol-dependent Bell-test benchmarks in quantum memory protocols relevant to quantum networks and long-distance quantum communication.


\begin{acknowledgments}
	
We sincerely thank Zi-Yu Liu and Jiun-Shiuan Shiu for valuable discussions and constructive feedback on the manuscript. This work was supported by the National Science and Technology Council of Taiwan under Grants No. 112-2112-M-006-034 and No. 113-2112-M-006-028. We also acknowledge support from the Center for Quantum Science and Technology within the framework of the Higher Education Sprout Project by the Ministry of Education in Taiwan.
	
\end{acknowledgments}

\vspace*{0.8\baselineskip}

\section*{Data Availability}

The data supporting the findings of this study are available from the corresponding author upon reasonable request.


\appendix

\section{EXPANSION OF MULTIMODE VACUUM OUTER PRODUCT} \label{sec:Appendix A}

The outer product of the multimode vacuum state, denoted as $|0\rangle = |0_{k_i}\ldots 0_{k_{i'}}\rangle$, is intrinsically related to the ladder operators of the optical field \cite{Vacuum Outer Product Property}. Its expansion can be expressed as follows:
\begin{align}
	\sum_{l=0}^{\infty}&\frac{(-1)^l}{l!}(\frac{2\pi}{L})^l\int_{k_1}\ldots\int_{k_l}\widetilde{a}_{k_l}^{\dagger}\ldots\widetilde{a}_{k_1}^{\dagger}\widetilde{a}_{k_1}\ldots\widetilde{a}_{k_l}dk_1\ldots dk_l\notag\\
	=&\sum_{m_{k_i}=0}^{\infty}\ldots\sum_{m_{k_{i'}}=0}^{\infty}\sum_{n_{k_j}=0}^{\infty}\ldots\sum_{n_{k_{j'}}=0}^{\infty}\sum_{l=0}^{\infty}\frac{(-1)^l}{l!}(\frac{2\pi}{L})^l|m_{k_i}\ldots m_{k_i'}\rangle\langle n_{k_j}\ldots n_{k_{j'}}|\notag\\
	&\times\int_{k_1}\ldots\int_{k_l}dk_1\ldots dk_l\langle m_{k_i}\ldots m_{k_i'}|\widetilde{a}_{k_l}^{\dagger}\ldots\widetilde{a}_{k_1}^{\dagger}\widetilde{a}_{k_1}\ldots\widetilde{a}_{k_l}|n_{k_j}\ldots n_{k_{j'}}\rangle\notag\\
	=&\sum_{m_{k_i}=0}^{\infty}\ldots\sum_{m_{k_{i'}}=0}^{\infty}\sum_{n_{k_j}=0}^{\infty}\ldots\sum_{n_{k_{j'}}=0}^{\infty}\sum_{l=0}^n\frac{(-1)^l}{l!}(\frac{2\pi}{L})^l|m_{k_i}\ldots m_{k_i'}\rangle\langle n_{k_j}\ldots n_{k_{j'}}|\langle m_{k_i}\ldots m_{k_i'}|\notag\\
	&\times\Big[\int_{k_l}\widetilde{n}_{k_l}dk_l\Big]\Big[\int_{k_{l-1}}\widetilde{n}_{k_{l-1}}dk_{l-1}-\frac{L}{2\pi}\Big]\ldots\Big[\int_{k_1}\widetilde{n}_{k_1}dk_1-\frac{L}{2\pi}(l-1)\Big]|n_{k_j}\ldots n_{k_{j'}}\rangle\notag\\
	=&\sum_{m_{k_i}=0}^{\infty}\ldots\sum_{m_{k_{i'}}=0}^{\infty}\sum_{l=0}^n\frac{(-1)^l}{l!}|m_{k_i}\ldots m_{k_i'}\rangle\langle m_{k_i}\ldots m_{k_i'}|n(n-1)\ldots(n-l+1)\notag\\
	=&\sum_{m_{k_i}\ldots m_{k_{i'}}}|m_{k_i}\ldots m_{k_i'}\rangle\langle m_{k_i}\ldots m_{k_i'}|\sum_{l=0}^n(-1)^l C^n_l\notag\\
	=&\sum_{m_{k_i}\ldots m_{k_{i'}}}\delta_{m_{k_i}+\ldots+m_{k_{1'}}, 0}|m_{k_i}\ldots m_{k_i'}\rangle\langle m_{k_i}\ldots m_{k_i'}|\notag\\
	=&|0_{k_i}\ldots 0_{k_{i'}}\rangle\langle 0_{k_i}\ldots 0_{k_{i'}}|, \label{eq:(57)}
\end{align}
where $n = m_{k_i} + \ldots + m_{k_{i'}}$, $\widetilde{n}_{k_i} = \widetilde{a}_{k_i}^\dagger \widetilde{a}_{k_i}$, and $C^n_l = \frac{n!}{l!(n-l)!}$ denotes the binomial coefficient. This expansion enables the outer product of two multimode number states to be expressed as a summation of products of ladder operators.


\section{MULTIMODE THEOREM} \label{sec:Appendix B}

The multimode field operator is required to satisfy the following commutation relations:
\begin{gather}
	[\hat{a}(z, t), \hat{a}^{\dagger}(z', t)]=L\delta(z-z'),\label{eq:(46)}\\
	[\widetilde{a}(k, t), \widetilde{a}^{\dagger}(k', t)]=\frac{L}{2\pi}\delta(k-k'),\label{eq:(47)}
\end{gather}
where the momentum-space field operator $\widetilde{a}(k, t)$ represents the Fourier transform of $\hat{a}(z, t)$. The multimode number basis in momentum space is defined as:
\begin{align}
	|n_{k_i}\ldots n_{k_{i'}}\rangle=|n_{k_i}{\rangle}_{k_i}\ldots|n_{k_{i'}}{\rangle}_{k_{i'}}=\otimes_{r=i}^{i'}|n_{k_r}{\rangle}_{k_r}.\label{eq:(48)}
\end{align}
Here, $n_{k_r}$ represents the number of photons in the momentum mode $k_r$, where $r \in \{i, \ldots, i'\}$. The multimode single-photon Fock state can then be expressed either in the spatial domain or the momentum domain as follows:
\begin{gather}
	|1; \phi(z, t)\rangle=\left[\frac{1}{\sqrt{L}}\int_{-\infty}^{\infty}\phi(z, t)\hat{a}^{\dagger}(z, t)dz\right ]|0\rangle=\frac{1}{\sqrt{L}}\int_{-\infty}^{\infty}\phi(z, t)|1_{z, t}\rangle dz,\label{eq:(49)}\\
	|1; \phi'(k, t)\rangle=\left[\sqrt{\frac{2\pi}{L}}\int_{-\infty}^{\infty}\phi'(k, t)\widetilde{a}^{\dagger}(k, t)dk\right ]|0\rangle=\sqrt{\frac{2\pi}{L}}\int_{-\infty}^{\infty}\phi'(k, t)|1_{k, t}\rangle dk,\label{eq:(50)}
\end{gather}
where the distribution functions are normalized ($\int_{-\infty}^{\infty}|\phi(z, t)|^2 dz = \int_{-\infty}^{\infty}|\phi'(k, t)|^2 dk = 1$) and related by $\phi'(k, t) = \frac{1}{\sqrt{2\pi}} \int_{-\infty}^{\infty} \phi(z, t)e^{ikz}dz$. The representation $|1_{z(k), t}\rangle = |0_{z_i(k_i)} \ldots 1_{z(k)} \ldots$ $0_{z_{i'}(k_{i'})}{\rangle}_t$ denotes a state with a single excitation at position $z$ (or momentum $k$) at time $t$, with no excitations elsewhere. Here, the number operators corresponding to the photon count within a specific spatial or momentum range are defined as follows:
\begin{gather}
	\hat{n}_{z_1, z_2}(t)=\int_{z_1}^{z_2}\frac{1}{L}\hat{a}^{\dagger}(z, t)\hat{a}(z, t)dz,\label{eq:(51)}\\
	\hat{n}_{k_1, k_2}(t)=\int_{k_1}^{k_2}\frac{2\pi}{L}\widetilde{a}^{\dagger}(k, t)\widetilde{a}(k, t)dk.\label{eq:(52)}
\end{gather}
The average photon number of the single-photon Fock state is $n_{|1\rangle} = \int_{-\infty}^{\infty} \frac{1}{L} \langle \hat{a}^\dagger(z, t)\hat{a}(z, t) \rangle dz = \int_{-\infty}^{\infty} \frac{2\pi}{L} \langle \widetilde{a}^\dagger(k, t)\widetilde{a}(k, t) \rangle dk = 1$. The lowering operator and inner product properties of the Fock state $|n_{k, t}\rangle$ are derived using mathematical induction \cite{Mathematical Induction}, as shown below:
\begin{gather}
	\widetilde{a}(k', t)|n_{k, t}\rangle=\frac{L}{2\pi}\delta(k-k')\sqrt{n_k}|(n-1)_{k, t}\rangle,\label{eq:(53)}\\
	\langle m_{k', t}|n_{k, t}\rangle=\frac{L}{2\pi}\delta(k-k')\delta_{m_{k'}, n_k},\label{eq:(54)}
\end{gather}
where Eq. (\ref{eq:(53)}) applies to $n_k > 0$ (with $\widetilde{a}(k', t)|0_{k, t}\rangle = 0$), and Eq. (\ref{eq:(54)}) applies to $n_k > 0 \vee m_{k'} > 0$ (where $\langle 0_{k', t}|0_{k, t}\rangle = 1$). Using Eq. (\ref{eq:(54)}), the inner product between the multimode single-photon Fock state and the multimode number basis is derived as:
\begin{align}
	\langle n_{k_i}\ldots n_{k_{i'}}|1; \phi'(k, t)\rangle=\sqrt{\frac{L}{2\pi}}\delta_{n_{k_i}+\ldots+n_{k_{i'}}, 1}\phi'(k_j, t)|_{n_{k_j}=1},\label{eq:(55)}
\end{align}
and the multimode single-photon Fock state in momentum space can be expanded as:
\begin{align}
	|1; \phi'(k, t)\rangle=\sqrt{\frac{L}{2\pi}}\sum_{n_{k_i}\ldots n_{k_{i'}}=10\ldots 0}^{0\ldots 01}\phi'(k_j, t)|1_{k_j, t}\rangle|_{n_{k_j}=1}.\label{eq:(56)}
\end{align}



\end{document}